\documentclass[12pt]{article}
\usepackage{jheppub}
\usepackage{amsmath}
\usepackage{amssymb}
\usepackage{slashed}
\usepackage{epstopdf}
\usepackage{asymptote}
\usepackage{courier}
\usepackage{empheq}
\newcommand{\dd}{\mathrm{d}}
\newcommand{\e}{\mathrm{e}}
\newcommand{\bea}{\begin{eqnarray}}
\newcommand{\eea}{\end{eqnarray}}
\newcommand{\ba}{\begin{align}}
\newcommand{\ea}{\end{align}}
\newcommand{\be}{\begin{eqnarray}}
\newcommand{\ee}{\end{eqnarray}}

\usepackage{mdframed}
\usepackage{dsfont}
\title{Emergent Lifshitz scaling from ${\cal N}=4$ SYM with supersymmetric heavy-quark density}
\preprint{ICCUB-14-070}
\author[a]{Anton F. Faedo,} 
\author[b]{Benjo Fraser,}
\author[c]{and S. Prem Kumar\,}
\affiliation[a]{Departament de F\'\i sica Fonamental and Institut de Ci\`encies del Cosmos, \\ Universitat de Barcelona,Mart\'\i\  i Franqu\`es 1, ES-08028, Barcelona, Spain.}
\affiliation[b]{Department of Physics, University of Athens,\\
15771 Athens, Greece.}
\affiliation[c]{ 
 Department of Physics,
 Swansea University, \\
Singleton Park,
 Swansea, SA2 8PP, U.K.  
}
\emailAdd{afaedo@ffn.ub.es}
\emailAdd{bjfraser@phys.uoa.gr}
\emailAdd{s.p.kumar@swansea.ac.uk}

\begin{document}

\abstract{We consider supersymmetric configurations in Type IIB supergravity obtained by the beackreaction of fundamental strings ending on a stack of D3-branes and smeared uniformly in the three spatial directions along the D3-branes. These automatically include a distribution of D5-brane baryon vertices necessary to soak up string charge.
The backgrounds are static, preserving eight supersymmetries, an $SO(5)$ global symmetry and symmetry under spatial translations and rotations.  We obtain the most general  BPS configurations consistent with the symmetries.  We show that the solutions to the Type IIB field equations are completely specified by a single function (the dilaton) satisfying a Poisson-like equation in two dimensions. We further find  that the equation admits a class of solutions displaying Lifshitz-like scaling with dynamical critical exponent $z=7$.  The equations also admit an asymptotically AdS$_5\times{\rm S}^5$ solution deformed by the presence of backreacted string sources that yield a uniform density of heavy quarks in ${\cal N}=4$ SYM.}

\maketitle

\newpage
\section{Introduction and Summary}
Theories with spatially isotropic (nonrelativistic) scale invariance emerge as descriptions at quantum criticality \cite{grinstein, hornreich, sachdev, fradkin} of various condensed matter systems. Such fixed point theories  display a ``dynamical'' or Lifshitz scale invariance under the transformation,
\be
t\, \to\, \lambda^z\, t\,,\qquad\qquad \vec x\, \to\, \lambda\, \vec x\,,\qquad\qquad z\,\neq\, 1\,,
\ee
which acts differently on the spatial ($\vec x$) and temporal ($t$) coordinates. Theories at Lifshitz points and at strong coupling are particularly interesting as they arise in the context of strongly correlated electron systems and  models of high-$T_c$ superconductors.
The holographic duality between (large-$N$) quantum field theories (QFTs) and gravity/string theory \cite{maldacena, witten} has provided a natural setting for exploring properties of strongly interacting QFTs at Lifshitz points \cite{Kachru:2008yh,Taylor:2008tg}, which can be further extended to theories with Lifshitz-like scaling accompanied by hyperscaling violation \cite{Ogawa:2011bz,Huijse:2011ef,Dong:2012se}.

In this paper we will show that Lifshitz scaling arises in an interesting and unusual fashion in a family of supersymmetric ($\tfrac{1}{4}$-BPS) solutions within type IIB supergravity. 
The backgrounds in question have a natural interpretation as  long distance descriptions of a state in ${\cal N}=4$ supersymmetric Yang-Mills (SYM) theory at large-$N$ and strong coupling, with a spatially homogeneous distribution of static quark impurities. The configurations correspond to backreacted geometries of the intersections of D3-branes, F-strings (heavy quarks) and D5-branes (baryon vertices) \cite{Witten:1998xy}. The main result of this paper is the derivation of the most general conditions to be satisfied by static $\tfrac{1}{4}$-BPS configurations in IIB supergravity preserving $ISO(3)\times SO(5)$ symmetry\footnote{$ISO(3)$ is the symmetry group, including translations and rotations, of the three spatial dimensions of the gauge theory along which quark density is uniformly distributed.}, which includes solutions discussed in \cite{Faedo:2013aoa}, but also new ones as we describe below.

Our work is motivated by the goal of eventually obtaining holographic models suitable for understanding high density physics in QCD-like theories and unravelling the ``condensed matter physics of QCD'' \cite{Rajagopal:2000wf, Fukushima:2010bq, CasalderreySolana:2011us}  (albeit within holographic toy models). In order to make progress towards this goal, it is necessary to understand the dynamics of quark flavours and how they influence or backreact on the   gluonic degrees of freedom at strong coupling. There are two reasons for this: The first reason is technical and a direct consequence of the large-$N$ 't Hooft limit that accompanies any classical holographic dual description of gauge theories. Since QCD has ``unquenched'' quarks, it is necessary to address the backreaction of quark flavours in the large-$N$ theory to model unquenched flavours. The second factor that necessitates inclusion of quark backreaction is intrinsic to physics at finite or high quark densities (in a deconfined phase) in the absence of temperature or any other comparable scale in the problem. In such a situation, when the quark density is not parametrically small, its backreaction on the large-$N$ vacuum will determine the ground state of the system. 

The necessity of incorporating flavour backreaction effects in holographic models at finite density (and low temperatures) has been emphasized in \cite{Hartnoll:2009ns,Bigazzi:2013jqa}. While both issues above should be addressed simultaneously in principle, it is useful and interesting to first understand the possible manifestations of backreacting quark density within a holographic setting. If the quark flavours are (sufficiently) massive, at low enough energies  we expect to be able to treat them as static objects. A state with a uniform density of these static quarks should, however, also be expected to backreact non-trivially on the glue degrees of freedom provided the quark density is $\sim {\cal O}(N^2)$ in the large-$N$ limit.

In gauge theory, a heavy static quark corresponds to a straight timelike Wilson line (e.g.\cite{Korchemsky:1991zp}). Therefore the state with a finite density of heavy-quarks can be viewed as the insertion of a distribution of Wilson line operators into the gauge theory. For densities scaling as $N^2$ in the large-$N$ limit, we expect a non-trivial ground state (saddle-point) to emerge. This idea was implemented in a non-supersymmetric fashion in \cite{Kumar:2012ui} in ${\cal N}=4$ SYM theory at large-$N$ and strong 't Hooft coupling. A heavy quark or straight Wilson line in ${\cal N}=4$ SYM corresponds to a macroscopic, infinite string stretching radially from the conformal boundary of  AdS$_5\times{\rm S}^5$ to the interior \cite{Maldacena:1998im, Rey:1998ik}. As is well known, this (BPS-)Wilson line also carries an orientational $SO(6)$ index associated to its location on the internal S$^5$. In \cite{Kumar:2012ui}, the distribution of static quarks was chosen to be both spatially uniform and 
$SO(6)$-symmetric, i.e. uniformly smeared around the S$^5$. The resulting non-supersymmetric background was shown to exhibit a flow from AdS$_5 \times {\rm S}^5$ to an IR geometry Lif$_5\times {\rm S}^5$
displaying Lifshitz scaling with dynamical critical exponent $z=7$. The scale invariance was found to be mildly broken by a logarithmically running dilaton\footnote{The existence of this scaling solution has also been noted in \cite{Azeyanagi:2009pr}.}. Since the ${\cal N}=4$ theory is scale invariant, there is no small parameter that controls the appearance of this IR scaling regime; any non-zero quark density leads to this (approximate) Lifshitz point. 

The non-trivial picture above leaves several open questions. Firstly, the significance and the origin of the numerical value of $z=7$ was {\it a priori} not understood. Second, given that the $SO(6)$-symmetric configuration is non-supersymmetric, the stability of the IR scaling solution was not established. Finally, although it is fairly clear that the picture should apply for dynamical massive quark flavours at low enough energy scales, its relevance for the large-$N$ theory with massless quark flavours requires further clarification\footnote{We would like to thank David Mateos and Javier Tarrio for enlightening discussions on this issue.}.

In this paper, following on from an earlier publication \cite{Faedo:2013aoa}, we derive the general BPS configurations describing  supersymmetric backgrounds preserving eight supercharges in type IIB supergravity, generated by smeared strings  intersecting with or ending on\footnote{Here we would like to make a distinction between the two situations: a string ending on a brane is semi-infinite and terminates at its endpoint on the brane, whilst a string intersecting a D-brane is of infinite extent and pierces through the D-brane.} a stack of D3-branes. In the latter case when semi-infinite strings end on branes, a non-vanishing density of baryon vertices or D5-branes is automatically induced. These appear and are necessary in order to soak up the string charge. The configurations described by our equations preserve an $SO(5)$ subgroup of R-symmetry. The intersecting brane solutions with vanishing D5-charge were already explored in \cite{Faedo:2013aoa} and they are interpreted as quark-antiquark pairs with antipodal $SO(6)$ orientation smeared uniformly in the gauge theory. The analysis in the present work allows us to further explore the situation where the strings end on the D3-branes and act as  a source for non-zero quark density with all the quarks aligned with the same internal orientation (preserving an $SO(5)$ internal symmetry).

The $\tfrac{1}{4}$-BPS configurations we find in this paper are determined by a single function, namely the dilaton, which satisfies a Poisson-like equation in two dimensions. It is somewhat remarkable then that this Poisson equation admits a class of solutions that exhibit Lifshitz scaling with $z=7$ and a logarithmically running dilaton. Together with \cite{Kumar:2012ui} and the F1-D3 intersection of \cite{Faedo:2013aoa}, this provides the third distinct instance of backreacted quark impurities in ${\cal N}=4$ SYM yielding identical scaling behaviour, independent of the global symmetries or supersymmetries of the configuration. This lends strong support to the physical picture found in \cite{Kumar:2012ui} -- that the static quark impurities trigger an RG flow at strong coupling in ${\cal N}=4$ SYM to the long-wavelength Lifshitz scaling regime with $z=7$. The F1-D3 intersections examined in \cite{Faedo:2013aoa} suggested that this scaling was a specific instance of a general dynamical critical exponent for F1-Dp intersections with $z = \frac{16-3p}{4-p}$. Indeed, a recent extensive and systematic study of smeared string configurations in Dp-brane theories with $p<6$ has revealed this scaling accompanied by 
hyperscaling violating behaviour \cite{Faedo:2014ana}. We will return to the potential significance of these results for addressing the larger questions that formed the motivation for this work, at the end of this paper. 

In section 2 we present the ans\"atze and the broad categories of the BPS configurations for the setup described above. The detailed analysis of BPS conditions and the derivation of the equation to be satisfied, is presented in the appendices. In section 3 we obtain some solutions to the Poisson-like equation that determines the supersymmetric backgrounds of interest. We summarise our results and discuss future directions in section 4.

\section{The setup and ans\"atze}

The backgrounds we are interested in result from the backreaction of mutually BPS Wilson lines, represented in the brane picture by parallel strings ending on or intersecting a stack of $N$ D3-branes. The worldine ${\cal C}$ of a static heavy quark associated to a string endpoint on the D3-branes is a straight timelike line, corresponding to the Maldacena-Wilson line which naturally incorporates a coupling with the six real scalars of the ${\cal N}=4$ theory,
\begin{align}
W_{\cal R}[\mathcal{C}]\,=\, \mathrm{Tr}_{\cal R}\mathcal{P}\exp\, \int_{\cal C}(i\,\dot{x}^{\mu} A_{\mu} + n^i\phi_i)\,\dd s\,.
\end{align}
Here $\phi_i$, $i=1,\cdots ,6$ are the scalars in the $\mathcal{N}=4$ vector multiplet, and $n$ is a constant unit six-vector.  Note that one also needs to specify the  representation $\mathcal{R}$ for the Wilson line. A single such Wilson line preserves the following set of symmetries: An $SO(5)$ subgroup of the $SO(6)$ R-symmetry group of ${\cal N}=4$ SYM, 16 of the 32 supercharges, a spatial rotational $SO(3)$ symmetry, and an $SL(2,\mathbb{R})$ subgroup of the full conformal group (generated by time translation, dilatation, and timelike special conformal transformations).  Depending on the representation ${\cal R}$ and for small enough representations, the holographic dual involves either fundamental F-string probes or wrapped D3- and D5-brane probes carrying string charge \cite{Drukker:2005kx, yamaguchi, Gomis:2006sb, Hartnoll:2006is}. For large representations with 
${\rm dim}[\mathcal{R}]\sim{\cal O}(N^2)$, the D-branes/Wilson loops can backreact to create a smooth supergravity geometry. Such backreacted ``bubbling'' geometries have been constructed in \cite{yamaguchi1,lunin,deg} for a single Wilson line. 

For the configurations that we are interested in, the number density of mutually BPS heavy quarks scales as $N^2$ and we replace them by a smeared uniform spatial distribution so as to restore translation invariance. However, in doing so we give up dilatation invariance, since a scaling of the spatial coordinates would change the smearing density. Thus the symmetries of our static setup are $\mathbb{R}_t\times ISO(3)\times SO(5)$, where $\mathbb{R}_t$ represents time translations. Such configurations preserve one quarter of the original supersymmetry or eight real supercharges.

In order to find the appropriate supersymmetric backgrounds, our strategy  is first to solve for the vanishing of the type IIB  supergravity supersymmetry variations, and then to check which equations of motion remain to be solved. An elegant method for doing this, pioneered in papers such as \cite{gauntlett11d,gmsw,gutowski}, is to work with all possible \emph{bilinears} of the Killing spinor, using the BPS equations to deduce the equivalent algebraic and differential equations governing the system. The details of our approach are presented in the appendix. 

\subsection{Supergravity ansatz}
Based on the symmetries preserved by the putative backgrounds, the ansatz for the Einstein frame metric is,
\begin{align}
\dd s_{\mathrm{Einstein}}^2 \,=\, \e^{2A} \dd x^i \dd x^i\, +\, \e^{2B} \dd\Omega_4^2\, +\, g_{\mu\nu} \dd x^{\mu} \dd x^{\nu}\,,\label{metric}
\end{align}
which describes an Euclidean three-plane and a four-sphere, with metric $\dd\Omega_4^2$, both fibered over a base Lorentzian (2+1)-dimensional manifold $\mathcal{M}_3$ which itself has a metric $g_{\mu\nu}$. This ansatz is invariant under the rotational and translational symmetries $ISO(3)$ of the $\mathbb{R}^3$, as well as the $SO(5)$ isometry group of the $S^4$. 

For the fluxes we choose the most general ans\"atze consistent with the same symmetries. Denoting the volume form of the manifold $X$ as $vol_X$, our ans\"atze for the fluxes read
\begin{align}
&F_5 \,=\, \mathcal{F}_2 \wedge vol_{\mathbb{R}^3} \,+\, \dd f \wedge vol_{S^4}\,,\\[2mm]
&G_3 \,=\, g \, vol_{\mathbb{R}^3}\, +\, h \, vol_{\mathcal{M}_3}\,,\nonumber\\[2mm]
&P\,=\, P (x^{\mu})\,,\qquad P_i\,=\,P_a\,=\,0\,,\nonumber
\end{align}
where $f \in \mathbb{R}$ and $g,h \in \mathbb{C}$ are functions on $\mathcal{M}_3$, while $\mathcal{F}_2$ is a real two-form on $\mathcal{M}_3$. The complex one-form $P$ contains the axio-dilaton \eqref{axiodilaton} and has components only along $\mathcal{M}_3$. The complex three-form $G_3$ encodes both the RR and NS three-forms. For the specific case of a vanishing axion it is given by $G_3 = \e^{-\phi/2} H_3 + i\, \e^{\phi/2} F_3$. Furthermore, ten dimensional self-duality of $F_5$ implies that
\begin{align}
\mathcal{F}_2 =  \e^{(3A-4B)} \ast_3\dd f
\end{align}
where $\ast_3$ is the Hodge star on $\mathcal{M}_3$. 

Upon substituting this ansatz into the fermionic variations of type IIB we find the most general supersymmetric configuration preserving (at least) 1/4 of the supercharges. The detailed analysis leading to this is  presented in appendices \ref{app:bpseq} and \ref{app:b}. In a particular S-duality frame in which the axion is vanishing, the supergravity fields take the following form
\bea
&&\dd s^2\, =\, -\, \e^{2(A+\phi)} \dd t^2\,  +\,\e^{2A} \dd x^i \dd x^i \,+\, \e^{-2A} \left[\e^{-\phi} \left(\dd y^2 \,+\, y^2 \dd\Omega_4^2\right)\,+\, \e^{\phi} \dd x^2 \right]\label{bpsmetric}\,,\\\nonumber\\
&& F_5\, =\, \frac{y^4}{4} (1+\ast)\, \left[- \partial_x (\e^{-4A-3\phi})  \dd y+\partial_y(\e^{-4A-\phi}) \dd x \right]\wedge vol_{S^4}\,,\\\nonumber\\
&&H_3 \,=\, \partial_y (\e^{2\phi})\,\dd t \wedge \dd y \wedge \dd x\,,\qquad \quad F_3 = -2\, \e^{4A}\partial_x (\e^{-\phi})\,  \dd x^1\wedge\dd x^2\wedge\dd x^3\label{3formfluxes}\,,
\eea
where $t$, $x$ and $y$ are coordinates on $\mathcal{M}_3$. Given that we are only interested in static configurations, the warp factor $A$ and the dilaton $\phi$ are only functions of $x$ and $y$. Note that we have solved for the warp factor $B$ appearing in \eqref{metric} in terms of $A$ and $\phi$. The SUSY projection conditions on the ten-dimensional complex spinor $\epsilon$ are
\be
\Gamma^{\hat{t}\,\hat{x}}\, D^{-1}\,\epsilon^* &\,=\,\epsilon\,\qquad\qquad
i\,\Gamma^{\hat{t}\,\hat{x}^1\hat{x}^2\hat{x}^3}\,\epsilon \,=\,\epsilon\,,
\eea
where $D$ is the complex conjugation matrix and hatted coordinates denote flat indices\footnote{The spinor $\epsilon$ takes the form $\epsilon=\e^{(A+\phi)/2}\epsilon_0$, where $\epsilon_0$ is a constant spinor satisfying the projections.}.

From the forms of the fluxes switched on, and the analysis presented in the appendix, it is clear that the supergravity fields are sourced by D3-, D5-branes and fundamental strings. 
We look for solutions which have at most localized (delta function) sources for D3 and D5 branes and fundamental strings. As explained in appendix \ref{eom} our equations can be easily extended to include smeared source distributions in the $x$-$y$ plane, but we will not explore these in this paper.
 Inspection of the equations of motion and the fluxes \eqref{3formfluxes} shows that the strings and branes in the system must have the following orientations
\begin{align}
\begin{array}{ccccccc}
&\quad x^1\quad&x^2\quad&x^3\quad&y\quad&x\quad&S^4\\[2mm]
{\rm F}1&\quad\cdot\quad&\cdot\quad&\cdot\quad&\cdot\quad&\times\quad&\cdot\\[2mm]
{\rm D}3&\quad\times\quad&\times\quad&\times\quad&\cdot\quad&\cdot\quad&\cdot\\[2mm]
{\rm D}5&\quad\cdot\quad&\cdot\quad&\cdot\quad&\times\quad&\cdot\quad&\times
\end{array}
\end{align}
The kappa symmetry conditions for these brane orientations are consistent with the SUSY projectors. 

The functions $A$ and $\phi$ are not arbitrary, but are determined by the Einstein equations and equations of motion for the fluxes.  We find that the solutions to the system of equations fall into two distinct categories:
\begin{figure}
\begin{center}
\includegraphics[width=2.7in]{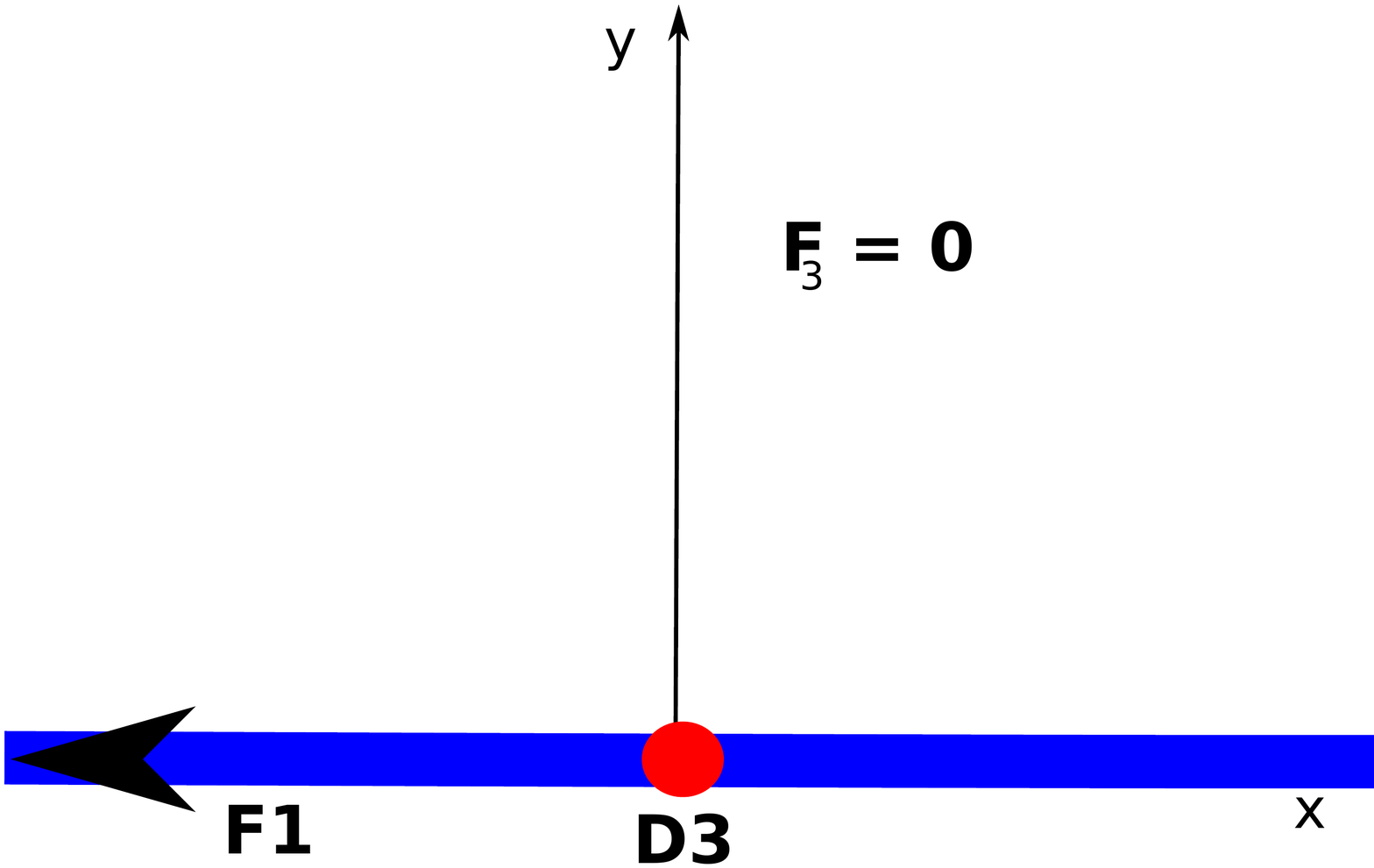}
\hspace{0.5in}
\includegraphics[width=2.7in]{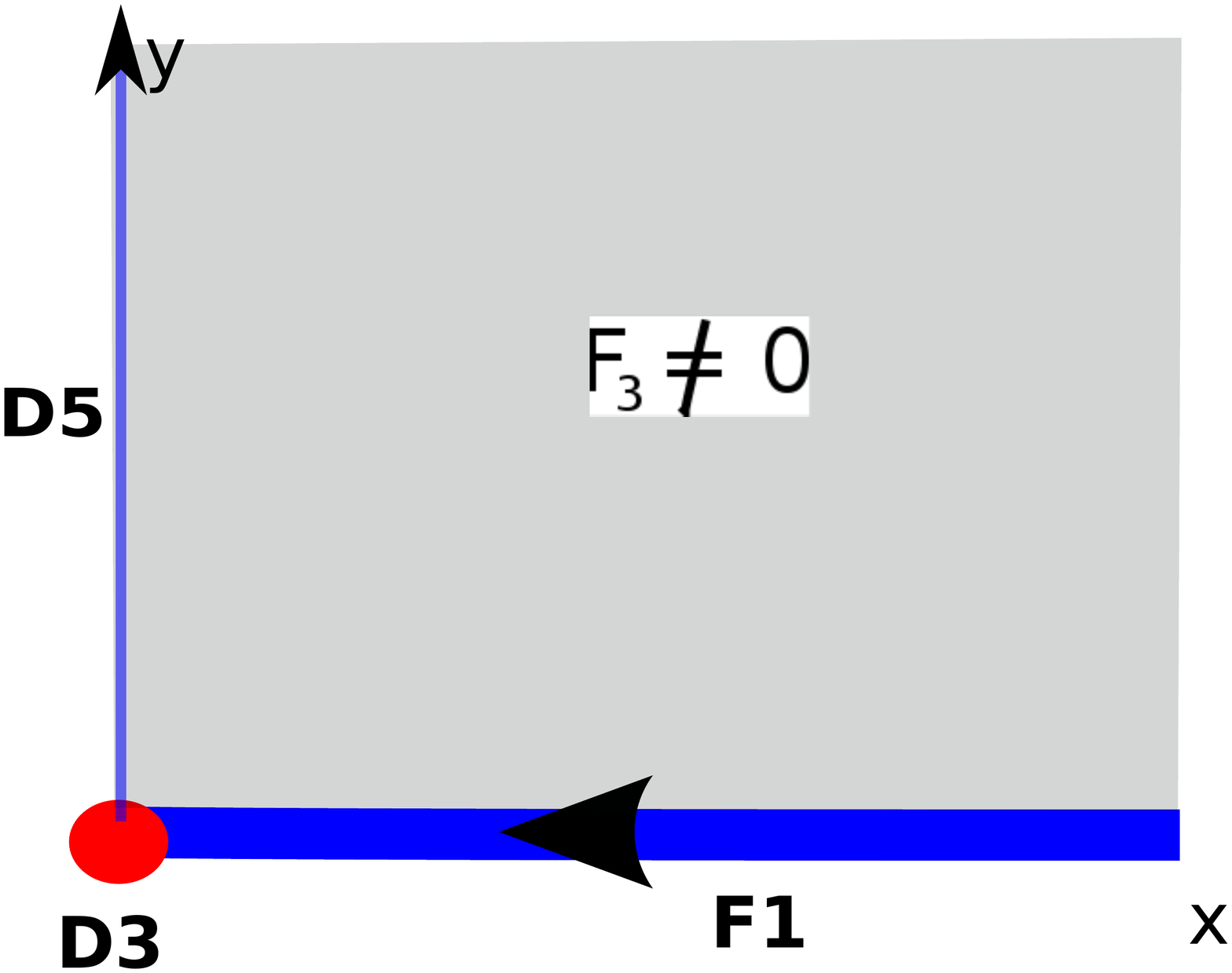}
\end{center}
\caption{\small {\bf Left}: Solutions with $F_3=0$ or intersecting F1-D3 configurations depicted on the $x-y$ plane with D3-branes at the origin and F-strings on the $x$-axis piercing through them.
{\bf Right}: Brane picture of configurations with $F_3\neq 0$ correspond to semi-infinite strings ending on the branes. Such solutions are also endowed with D5-brane baryon vertices smeared along the gauge theory directions with the D5-branes wrapping S$^4$ and extending along the $y$-axis. 
}
\label{fig1}
\end{figure}

\begin{itemize}
\item{{\bf Category I} ($F_3=0$): These solutions have a dilaton independent of $x$, i.e. $\partial_x\phi \,=0\,$, which immediately implies $F_3=0$, as can be seen  from eq.\eqref{3formfluxes}. Such backgrounds do not have 
D5-brane sources associated to them and are backreacted descriptions of  supersymmetric D3-F1 intersections. The D-brane picture is indicated in Fig.\eqref{fig1} and corresponds to a bundle of parallel infinite strings piercing/intersecting the D3-branes. The entire family of such solutions was explored in earlier work \cite{Faedo:2013aoa} and includes the  delocalised intersections studied in \cite{hsingh, deyroy}. Such configurations were shown to yield both Lifshitz and hyper scaling violating IR geometries. These included a partially localised intersection with $z=7$.}
\item{{\bf Category II} ($F_3\neq 0$): For the second category of solutions the dilaton depends both on $x$ and $y$ coordinates. This automatically leads to a non-vanishing three-form flux. From the orientation of this flux we infer that it can be associated to D5-branes wrapping S$^4\times {\mathbb R}_y$  and distributed uniformly along the spatial coordinates $x^{1,2,3}$. In particular,
\begin{equation}
F_3\,=\,Q_{\rm B}\, \dd x^1\wedge\dd x^2\wedge\dd x^3\,.
\end{equation}
for some constant $Q_{\rm B}$, proportional to the number of D5-branes. The precise normalization and its relation to the number density of strings can be inferred by comparison with \cite{Kumar:2012ui}. Consistency with (\ref{3formfluxes}) requires the warp factor to be specified in terms of the dilaton, that is,
\begin{equation}
\e^{-4A}\,=\,-\frac{2}{Q_{\rm B}}\,\partial_x \e^{-\phi}\label{Aphirelation}
\end{equation}
which remains the only function to be determined by the equations of motion. Notice that once this replacement is executed the limit $Q_{\rm B}\to0$ is not smooth and for that case we refer again to \cite{Faedo:2013aoa}. Using the intermediate results quoted in the appendices, it is easy to compute the Maxwell charges associated to the different branes and strings in the system:
\begin{eqnarray}
Q_{\rm D5}&=&\int_{\Sigma_3}F_3\,=\,Q_{\rm B}\,\,vol\left(\mathbb{R}^3\right)\,,\nonumber\\[2mm]
Q_{\rm D3}&=&\int_{\Sigma_5}4\,F_5\,=\,4\,f\,vol\left(S^4\right)\,=\,-\frac{1}{Q_{\rm B}}y^4\partial_y(\e^{-2\phi})
\,vol\left(S^4\right)\nonumber\\[2mm]
Q_{\rm F1}&=&\int_{\Sigma_7}
e^{-\phi}\ast H_3
\,=\,y^4\partial_y(\e^{-2\phi})
\,vol\left(\mathbb{R}^3\right)vol\left(S^4\right)\label{F1charge}
\end{eqnarray}
As usual, the Maxwell charges are not gauge invariant (not topological). Note also that they are not constants and can depend on the radial coordinates $x$ and $y$. Nevertheless, the expected relation between the number of D3-branes, the number of fundamental strings and the number of baryon vertices, which is of course gauge invariant, is verified
\begin{equation}\label{chargesrel}
Q_{\rm F1}\,=\,-Q_{\rm D5}\,Q_{\rm D3}\,.
\end{equation}

}
\end{itemize}

\section{Poisson-like equation and some solutions}

The endpoint of the analysis of the BPS configurations and the ensuing type IIB field equations is that the whole system is governed by a non-linear Poisson-like equation
\begin{align}
\boxed{\frac{1}{y^4}\partial_y (y^4\partial_y\e^{-2\phi})\,+\,\frac{1}{2}\,\partial_x^2\e^{-4\phi}\,=\,\rho(x,y)\,,}\label{poissoneq}
\end{align}
where we have allowed for a source term on the right hand side in keeping with the discussion in appendix \ref{sourcing}. In this paper we will not have explicit smeared sources on the $x$-$y$ plane. All branes and strings will be taken to be localized in these coordinates (consistent with the depiction in Fig.\eqref{fig1}), so that we can set $\rho(x,y)=0$. Recall, for instance, that the strings are located on the $x$-axis, corresponding to a delta-function source for the Poisson equation.

In general the BPS configurations solve the equations of motion as well as the Einstein and dilaton equations with source terms for the strings when we identify the smearing form $\Omega_8$ for the strings as $\Omega_8=-y^4\rho(x,y)\,\dd y\wedge\dd x^1\wedge\dd x^2\wedge\dd x^3\wedge vol_{S^4}$ (see appendix \ref{sourcing}).

The Poisson equation for the dilaton  is similar in spirit to the Toda equation which appears in the analysis of \cite{maldacenagaiotto, llm}. In that case there was an implicit variable change which mapped the problem into a linear electrostatics problem. In the present case we are not aware of such a simplification and are confronted with a nonlinear partial differential equation.

\subsection{Scaling solution with $z=7$}

We now observe that the homogeneous version of Eq.\eqref{poissoneq} possesses an interesting and physically relevant family of solutions. If we take
\be
\e^{-2\phi}\,=\,\frac{Q_1}{y^3}\Psi(p)\,,\qquad p\equiv
\frac{x^2 y}{Q_1}\,.\label{scalingdilaton}
\ee
where $Q_1$ is some constant, then the dependence on $y$ simply factors out of Eq.\eqref{poissoneq} and we obtain a non-linear ordinary differential equation for $\Psi(p)$: 
\begin{align}
p \,(4 \Psi\,+\,p)\, \Psi''\,+\,4\, p\, \Psi^{\prime\,2}\,+\,2 (\Psi-p) \,\Psi'\,=\,0\,.\label{scalingeq}
\end{align}
For any $\Psi(p)$ it is easily seen (using \eqref{bpsmetric} and \eqref{Aphirelation}) that the resulting metric is 
\bea
\dd s^2\,=&&\left(-\Psi'\Psi^{\frac{3}{2}}p^{\frac{1}{2}}\right)^{-\frac{1}{2}}\left[-\,y^{7/2}\,\dd t^2\,+\, y^{1/2}\Psi\,\dd x^i\dd x^i\,+\,\left(\frac{\dd y^2}{y^2}\,+\,\dd\Omega^2_4\right)\,(-\Psi')\Psi\,p^{\frac{1}{2}}\sqrt{\tfrac{2Q}{Q_{\rm B}}}\right.\nonumber\\\nonumber\\
&&\left.+\,y\,\dd x^2\,(-\Psi' p^{\frac{1}{4}})\sqrt{\tfrac{2}{Q\,Q_{\rm B}}}\,\right]\,.\label{z=7}
\eea
where some numerical and other constants have been absorbed into rescalings of the coordinates $t$ and $x^{i}$. 
 We have also taken the constant charge densities $Q_1$ and $Q_{\rm B}$ to be positive so that it is necessary for $\Psi'(p)$ to be negative definite for a sensible solution. This metric is invariant under the transformations:
\bea
t\mapsto \lambda^7 t\,,\qquad x^i\mapsto \lambda x^i\,,\qquad x\mapsto \lambda^2\, x\,,\qquad y\mapsto \lambda^{-4} y\,,
\eea
where $y$ plays the role of the standard radial coordinate. It is quite remarkable that a similar realization of the $z=7$ scaling was also found in the intersecting brane setup of \cite{Faedo:2013aoa} with vanishing $F_3$, through the dependence on the variables $p$ and $y$ as defined above. Note also that the equations of motion that determine the background for vanishing $F_3$ are effectively linear, so the actual solutions differ significantly from what we are discussing here. Since the dilaton is also scale dependent as in \eqref{scalingdilaton}, the background does not display exact scale invariance, precisely as found in the solutions of \cite{Kumar:2012ui, Faedo:2013aoa}.
The manner in which the scaling solution is realized is unusual because the background contains two radial coordinates $x$ and $y$, or equivalently, the pair  $p$ and $y$, accompanied by an S$^4$ factor.
 
We can now be more specific about the function $\Psi(p)$. For small $y$, as the $x$-axis is approached (we can do this by taking the limit $y\to 0$ first followed  by $p\to 0$) the dilaton \eqref{scalingdilaton} should be such that it yields the number density of fundamental strings as indicated by the Maxwell charge \eqref{F1charge} when $y\to 0$. Note however that this number does not correspond to the number density of strings in the microsopic setup or to the number density of heavy-quarks in the boundary gauge theory. This is because of the presence of a non-zero $F_3$ and $F_5$ in the system\footnote{In contrast, when $F_3=0$ as in \cite{Faedo:2013aoa} the string charge has no such ambiguity.}. Indeed, the number density of quarks in the gauge theory is determined by $Q_{\rm B}$, the baryon number density. Requiring $\Psi(p)$ to approach a finite value, normalized to unity, in the limit $y\to 0$ we find 
\begin{align}
\Psi (p) = 1\,+\,s_0\, p^{1/2}\, -\, \frac{s_0^2}{2}\, p +\frac{s_0}{24}\left(5\,+\,12s_0^2\right)\,p^{3/2}\,+\,\mathcal{O}\left(p^2\right)
\end{align}
where all higher terms in the expansion are determined by a single integration constant $s_0$. Keeping the leading term we recover the usual Lifshitz metric with $z=7$ associated with the string distribution, but the first correction is different from the one encountered in \cite{Faedo:2013aoa}
as the expansion in the present context involves half-integral powers of $p$.

\subsection{Asymptotically AdS$_5\times {\rm \bf S}^5$ solution}
It is an important consistency check to verify that  solutions to \eqref{poissoneq} yield a flow away from AdS$_5\times {\rm S}^5$ asymptotics in similar fashion to that encountered in \cite{Kumar:2012ui} and \cite{Faedo:2013aoa}. In particular, the AdS$_5\times {\rm S}^5$ vacuum, 
which has $F_3=0$ and thus falls into the first category of solutions, is obtained when
\be
\e^{-2\phi}\,=\,1\,,\qquad \e^{-4A}\,=\,\frac{1}{(x^2+y^2)^2}\,,\qquad
x\,=\,r\cos\theta\,,\qquad y\,=\,r\sin\theta\,.
\ee 
To extract the flow triggered by string sources, we linearize around the vanishing dilaton solution as
\be
 \e^{-2\phi}\,=\,1\,+\,\epsilon\, h_{(1)}\,+\,\mathcal{O}(\epsilon^2)\,. 
 \ee
The expansion parameter $\epsilon$ will eventually be related to the baryon vertex density $Q_{\rm B}$. We find that the first correction $h_{(1)}$ satisfies the $SO(5)$ symmetric Laplace equation on flat $\mathbb{R}^6$:
\begin{align}
\frac{1}{y^4}\,\partial_y (y^4\partial_y h_{(1)})\,+\,\partial_x^2h_{(1)}\,=\,0\,.
\end{align}
This Laplace equation possesses a large family of solutions  which takes the form of a sum over point charges,
\begin{align}
h_{(1)} = \frac{1}{y^2}\sum_i q_i \left[\frac{(x-x_i)}{(x-x_i)^2+y^2}+\frac{1}{y}\left(\arctan\frac{x-x_i}{y}+\frac{\pi}{2}\right)\right]\,.
\end{align}
Crucially, this family of solutions follows from the first order equation \eqref{Aphirelation}, upon using the linearized ansatz around AdS$_5\times {\rm S}^5$,
\be
\epsilon\,\partial_x h_{(1)}\,=\,- Q_{\rm B}\sum_i{q_i}\frac{1}{\left[(x-x_i)^2\,+\,y^2\right]^2}\,,
\ee
where, on the right hand side we have introduced a particular multicentre distribution of D3-branes corresponding to a Coulomb branch configuration of ${\cal N}=4$ SYM. The $q_i$ represent the fraction of the total number D3-branes placed at the position $x_i$, so that $\sum_i q_i =1$. This leads to the identification 
$Q_{\rm B}=-2\epsilon$, which implies that the $\epsilon$-expansion is equivalent to an expansion in $Q_{\rm B}$.
For simplicity we focus attention on the origin of the Coulomb branch so that $x_i=0$. Then we find the following asymptotic components of the spatial metric component in polar coordinates:
\begin{align}
h_{ii}\,=\,r^2
\,-\,\frac{Q_{\rm B}}{16}\frac{1}{r}\frac{1}{\sin^3\theta}\left(2(\pi-\theta)+\sin2\theta\right) + \mathcal{O}(r^{-4},Q_{\rm B}^2)\,.
\end{align}
This displays the $1/r$ potential term typical of backreacted string sources in AdS$_5$ \cite{Kumar:2012ui, Headrick:2007ca}. Similarly, the dilaton is
\begin{align}
\e^{-2\phi}=1-\frac{Q_{\rm B}}{2}\frac{2(\pi-\theta)+\sin 2\theta}{\sin^3\theta}\frac{1}{r^3}+\mathcal{O}(Q_{\rm B}^2)
\end{align}
and to this order in $Q_{\rm B}$ the five-sphere is undeformed with respect to the vacuum AdS$_5\times {\rm S}^5$. The dependence of these corrections on the polar angle $\theta$ of the  five-sphere is also consistent with the physical picture of the string sources being placed at one of the poles i.e at $\theta=0$. This is reflected in a singularity at this point in the metric and dilaton corrections. There is no such singularity at $\theta=\pi$.

\subsection{Smeared D5 solution}

There is another simple solution to the homogeneous non-linear Poisson equation, obtained by forcing each term in  the equation to vanish separately. It reads
\begin{equation}
\e^{-2\phi}\, =\, \left(1+2\,Q_{\rm B} \,x\right)^{\frac12}\left(1+\frac{Q_1}{y^3}\right)\equiv h_x^{\frac12}\,h_y\,.
\end{equation}
The metric can be written in terms of these warp factors as
\begin{equation}\label{intersection}
\dd s^2\,=\,h_x^{-\frac18}h_y^{\frac34}\left[h_y^{-2}\,\dd t^2+\left(\dd y^2+y^2\dd\Omega_4^2\right)\right]+h_x^{-\frac58}h_y^{-\frac14}\left[h_x\,\dd x^i\dd x^i+\dd x^2\right]\,,
\end{equation}
supported by the fluxes
\begin{eqnarray}
H_3&=&-h_x^{-\frac12}\,\dd h_y^{-1}\wedge\dd t\wedge\dd x\,,\nonumber\\[2mm]
F_3&=&Q_{\rm B}\,\dd x^1\wedge\dd x^2\wedge\dd x^3\,,\nonumber\\[2mm]
F_5&=&\frac{3}{4}\,\frac{Q_1}{Q_{\rm B}}\left(1+*\right)\dd h_x^{\frac12}\wedge vol_{S^4}\,.
\end{eqnarray} 
To interpret this solution it is useful to take the $Q_1\to0$ limit, i.e. $h_y\to1$, in which case there are neither strings nor D3-branes. In this limit, the time coordinate combines with $y$ and the four-sphere to form a six-dimensional Minkowski space in which D5-branes are extended. As can be seen through an harmonic superposition analysis, the solution describes the backreacted geometry due to D5-branes homogeneously distributed along the $x^i$ directions. Allowing for non-vanishing $Q_1$ includes smeared strings and the solution (\ref{intersection}) describes the intersection of the strings with the D5-branes. Five-form flux is induced so that condition (\ref{chargesrel}) is fulfilled.

\subsection{A different scaling solution}
 We have also found an exact scaling solution of the type in Eq.\eqref{scalingdilaton} with $\Psi(p)=\frac13p$. Changing to the radial variables $y=\rho^4$ and $x=\sqrt3 \,\sigma^{-2}$ this solution reads:
\begin{align}
\dd s^2\, &=L^2\, \left( -\rho^{10}\,\sigma^4 \dd t^2\, +\,  \rho^2\,\dd x^i\dd x^i\, +\, \frac{\dd\rho^2}{\sigma^2}\, +\, \frac 34 \frac{\rho^2}{\sigma^4}\, \dd\sigma^2\,+\, \frac{1}{16}\frac{\rho^2}{\sigma^2}\dd\Omega_4^2\right)\,,\\[2mm]
F_5\, &=\, \frac{L^6}{128\,Q_{\rm B}}\,(1+\ast)\,\dd \left( \frac{\rho^4}{\sigma^4}\right)\wedge vol_{S^4}\,,\\[2mm]
H_3\, &=\,4\sqrt{3} \,L^2\, \rho^7\sigma\, \dd t\wedge\dd\rho\wedge\dd\sigma\,,\\[2mm]
F_3\, &=\, Q_{\rm B}\, \dd x^1\wedge\dd x^2\wedge\dd x^3\,,\\[2mm]
\e^{\phi}\, &=\, \rho^4\sigma^2\,,
\end{align}
where $L=\frac{3^{1/4}}{2^{3/2}}Q_{\rm B}^{1/2}$. Surprisingly, the scaling properties of this metric are enhanced due to the interplay of both radii. Under a rescaling
\begin{equation}
t\mapsto \lambda^z\, t\,,\qquad x^i\mapsto \lambda\, x^i\,,\qquad \rho\mapsto \lambda^{\frac{3-z}{4}} \,\rho,\qquad \sigma\mapsto \lambda^{-1}\, \sigma,\qquad 
\end{equation}
the metric transforms as
\begin{equation}
\dd s\mapsto \lambda^{\frac{7-z}{4}}\,\dd s
\end{equation}
signaling certain  hyperscaling properties. Notice that the Lifshitz and hyperscaling coefficients are related but in principle arbitrary. We do not expect violations of energy conditions since the matter supporting the solution (smeared strings) does not have unphysical features.  

It is possible to absorb this rescaling of the metric into the radius through the parameter $Q_{\rm B}$. This would mean that under a dilatation we are only changing the baryon density, or equivalently the string density. If we do so, the only fields that transform are the dilaton, $H_3$ and $F_3$, but in such a way that the relation between charges (\ref{chargesrel}) is maintained. 

Interestingly, $F_5$ only depends on the combination $\rho^4/\sigma^4\sim x^2y = p$. An analogous argument to that in \cite{Faedo:2013aoa} tells us that the D3-branes are not localized in these solutions. We further note that the curvature in string frame is
\begin{equation}
\mathcal{R}_{\rm String}\,=\,-\frac{288}{L^2}\,\frac{\sigma}{\rho^4}\,.
\end{equation}
Therefore there is a curvature singularity situated all along the $x$ and $y$ axes.  The physical significance of these and other properties of this  solution remain to be understood.


%
%
%
%
%


\section{Conclusions}
The main motivation behind this work was to  understand whether the emergence of the IR $z=7$ scaling found in \cite{Kumar:2012ui} for the ${\cal N}=4$ theory with heavy quarks  could be reproduced within a supersymmetric setup. A significant outcome of our work is the derivation of the most general BPS configurations with eight supercharges in type IIB supergravity preserving $ISO(3)\times SO(5)$ global symmetry, showing that they are determined by the solutions to the two dimensional Poisson-like equation \eqref{poissoneq}. We found solutions to this equation with $z=7$ Lifshitz-like scaling, and also showed that the equation correctly captures the flow away from the asymptotically AdS$_5 \times {\rm S}^5$ regime. Obtaining the flow interpolating between these two limits requires numerical integration of the PDE \eqref{poissoneq} which is interesting work for the future.

It would be extremely interesting to understand if there is a general structure underlying the solutions of the Poisson equation \eqref{poissoneq} for this system, along the lines of the picture found in \cite{yamaguchi1, lunin, deg}. In particular, our equations, with suitable sources on the $x$-$y$ plane, should also be able to describe  distributions  of Wilson lines/quarks in more general representations. Such information should be contained in a general linearized analysis of the UV asymptotics around AdS$_5\times {\rm S}^5$ which we have not explored completely in this paper. It would be interesting to know whether different choices of impurity representations have any effect on the long wavelength/IR description of the system or if they all flow to the same Lifshitz-like scaling solution. 

The supersymmetric scaling solutions in this paper are singular due to the running of the dilaton. In order to make sense of such backgrounds it is important to have a non-extremal version of these solutions where the singular region is shielded behind a horizon and one may reliably speak about the scaling properties of physical quantities. This was easily achieved in the non-supersymmetric $SO(6)$ symmetric configuration of \cite{Kumar:2012ui}. The corresponding generalization to the $SO(5)$-symmetric setup of this paper is not obvious due to the presence of effectively two radial directions in the bulk solutions.

The results of this paper also lend support to the general idea of applying the smearing technique to understand holographic backreaction of quark flavours at finite density \cite{Faedo:2014ana, benini, carlos, bigazzi1, bigazzi2}, in that long-distance properties of systems may not be sensitive to the details of the smearing procedure itself. It is important, however, to understand in detail the embedding of the scaling solutions found in  \cite{Kumar:2012ui, Faedo:2014ana} within  the setup of backreacted and smeared flavoured holographic duals \cite{carlos}. This would open the way for understanding possible instabilities and their end-points that may lead to (colour) superconducting phases along the lines of the ideas presented in \cite{Hartnoll:2009ns, Chen:2009kx, obannon}

\acknowledgments 
We would like to thank Paolo Benincasa, John Estes, David Mateos, Andy O'Bannon, and Javier Tarrio for many discussions on several topics of immediate relevance to this work. SPK would also like to thank David Mateos and the Departament de F\'\i sica Fonamental and Institut de Ci\`encies del Cosmos, University of Barcelona, for providing a stimulating atmosphere in which this work could be completed.  AF acknowledges financial support from the grants MEC FPA2010-20807-C02-02, by CPAN CSD2007-00042 Consolider-Ingenio 2010 and by the ERC Starting Grant ÒHoloLHC-306605Ó.
The research of BF is implemented under the ``ARISTEIA'' action of the ``operational programme education and lifelong learning'' and is co-funded by the European Social Fund (ESF) and National Resources. 
SPK thanks the U.K. Science and Technology Facilities Council (STFC) under the grants ST/J000043/1 and ST/L000369/1, for financial support.

\appendix
\section*{Appendix}

\section{BPS equations}
\label{app:bpseq}

We use the conventions of \cite{deg}, which are stated explicitly in \cite{deg1}. Note that the normalization of $F_5$ differs from that usually used in string theory by a factor of 4.
Type IIB supergravity is written in terms of two field strengths $P$ and $Q$, and string theory in terms of the dilaton $\phi$ and axion $C^{(0)}$. Following e.g. \cite{gmsw}, we write the map between the supergravity and string theory variables:
\begin{align}
P\,=\,\frac 12 \dd\phi\,+\,\frac i2 \e^{\phi}\dd C^{(0)}\,,\qquad\qquad Q \,=\,-\frac 12\e^{\phi}\dd C^{(0)}\,.\label{axiodilaton}
\end{align}
The supergravity equations of motion have a local $U(1)$ invariance with associated gauge field $Q$. Each field has a definite charge $q$ under this $U(1)$: $\epsilon$ has $q=1/2$, $P$ has $q=2$ and $G_3$ has $q=1$. The field strengths have corresponding Bianchi identities written in terms of the $U(1)$-covariant derivative $\mathrm{D}\equiv \nabla -iqQ$
\begin{align}
\mathrm{D}P&=0\label{pbianchi}\\
\dd Q&=-i P\wedge P^*\label{qfieldstrength}
\end{align}
which are automatically satisfied when we use the map to string theory variables. 
This formulation comes from a gauge fixing of the version of the theory with an extra auxiliary scalar field, and the remnant of this is that each $SL(2,\mathbb{R})$ action is accompanied by a local $U(1)$ gauge transformation. This is the only way in which $SL(2,\mathbb{R})$ duality acts on the variables $G, P, Q$. 

Type IIB supergravity has 32 real supercharges parametrized by a complex chiral ten-dimensional spinor $\Gamma \epsilon = - \epsilon$. We begin by writing down the SUSY variations in the Einstein frame:
\begin{align}
\delta_{\epsilon} \lambda &= i (\Gamma\cdot P) \mathcal{B}^{-1} \epsilon^* - \frac{i}{24}(\Gamma\cdot G)\, \epsilon \label{dilatino}\\
\delta_{\epsilon} \psi_M &= \mathrm{D}_M \epsilon +\frac{i}{480} (\Gamma\cdot F) \Gamma_M \epsilon -\frac{1}{96} \left[ \Gamma_M (\Gamma \cdot G) + 2 (\Gamma \cdot G) \Gamma_M \right] \mathcal{B}^{-1} \epsilon^*\qquad .\label{gravitino}
\end{align}

The next step is to choose a basis of gamma matrices in ten dimensions. We choose:
$$
\begin{array}{ccccccccc}
\Gamma_i &=& \gamma_i &\otimes & \gamma_{S^4} &\otimes & \mathds{1} &\otimes & \sigma^1\\
 \Gamma^a &=& \mathds{1} &\otimes & \gamma^a &\otimes & \mathds{1} &\otimes & \sigma^1\\
\Gamma_{\mu} &=& \mathds{1} &\otimes & \mathds{1} &\otimes & \gamma^{\mu} &\otimes & \sigma^2
\end{array}
$$
\noindent where $\gamma_{S^4}\equiv+\gamma^6\gamma^7\gamma^8\gamma^9$ is the chirality matrix on $S^4$. The ten dimensional chirality matrix is $\Gamma = \mathds{1} \otimes \mathds{1} \otimes \mathds{1} \otimes \sigma^3$, so that the IIB chirality condition reduces to
\begin{align}
\sigma^3 \epsilon = - \epsilon
\label{chirality}
\end{align}

We complete the basis by specifying gamma matrices within each factor space:
$$
\begin{array}{cccccc}
i\, : \qquad &\gamma^3 = \sigma^1,\, &\gamma^4 = \sigma^2,\, &\gamma^5 = \sigma^3&&\\
a\, : \qquad &\gamma^6 = \sigma^1 \otimes \mathds{1},\, &\gamma^7 = \sigma^2 \otimes \mathds{1},\, &\gamma^8 = \sigma^3 \otimes \sigma^1,\, &\gamma^9 = \sigma^3 \otimes \sigma^2&(\Rightarrow\gamma_{S^4}=-\sigma^3\otimes\sigma^3)\\
\mu\, : \qquad &\gamma^0 = i \sigma^2,\, &\gamma^1 = \sigma^1,\, &\gamma^2 = \sigma^3&&
\end{array}
$$

\noindent Note that the basis for the gamma matrices on $\mathcal{M}_3$ is real (Majorana). The ten dimensional complex conjugation matrix $\mathcal{B}$, defined by $\{\mathcal{B}\, \Gamma^{\mu} \mathcal{B}^{-1} = (\Gamma^{\mu})^*, \, \mathcal{B}^* \mathcal{B} =\mathds{1}\}$, is now
\bea
\mathcal{B} &=& \sigma^2 \otimes (\sigma^2 \otimes \sigma^1) \otimes \mathds{1} \otimes \sigma^3\\
&=& b_3 \otimes b_4 \otimes \mathds{1} \otimes \sigma^3
\eea
\noindent where $b_3$ and $b_4$ are charge conjugation matrices in $\mathbb{R}^3$ and $S^4$ respectively:
$$
b_3 \gamma^i b_3^{-1} = - (\gamma^i)^* \qquad b_4 \gamma^a b_4^{-1} = - (\gamma^a)^*
$$

We plug our ansatz and our Clifford algebra basis into the IIB SUSY variations \eqref{dilatino}\eqref{gravitino}, and after a few pages of careful work we end up with the following set of BPS conditions on our ten dimensional complex spinor $\epsilon$:
\begin{align}
\slashed{P} \mathcal{B}^{-1} \epsilon^* - \frac{1}{4}(\e^{-3A} g \gamma_{S^4} - h) \epsilon &= 0\\
\frac 13\e^{-A}\gamma^i \tilde{\nabla}_i \epsilon+\frac{i}{2} \slashed{\partial} A  \gamma_{S^4} \epsilon + \frac{1}{2} \e^{-4B} \slashed{\partial} f  \epsilon - \frac{i}{16} (3 \e^{-3A} g + h \gamma_{S^4} ) \mathcal{B}^{-1} \epsilon^* &= 0\\
\e^{-B} \tilde{\nabla}_a \epsilon - \frac{i}{2} \gamma_a \slashed{\partial} B \epsilon + \frac{1}{2} \e^{-4B} \gamma_a \slashed{\partial} f \gamma_{S^4} \epsilon - \frac{i}{16} \gamma_a (\e^{-3A} g \gamma_{S^4} - h) \mathcal{B}^{-1} \epsilon^* &= 0\\
\mathrm{D}_{\mu}\epsilon + \frac{i}{2} \e^{-4B} \slashed{\partial} f \gamma_{S^4} \gamma_{\mu} \epsilon + \frac{1}{16} (\e^{-3A} g \gamma_{S^4} + 3 h)\gamma_{\mu} \mathcal{B}^{-1} \epsilon^* &= 0
\label{kse1}
\end{align}
where here and in the following $\mathrm{D}_{\mu}$ and $\nabla_{\mu}$ denote derivatives on $\mathcal{M}_3$, and $\tilde{\nabla}_{i,a}$ are derivatives on $\mathbb{R}^3$ and $S^4$ respectively. Also note that when an operator appears which naturally acts within only one Clifford subspace, it should be taken as the tensor product with the identity matrix in the other tensor factors. For example, by $\gamma_{S^4}\epsilon$ we mean $(\mathds{1}\otimes\gamma_{S^4}\otimes\mathds{1}\otimes\mathds{1})\epsilon$. 

To proceed we must make an ansatz for the form of the ten dimensional spinor $\epsilon$
\begin{align}
\epsilon = \eta^{\alpha} \otimes \chi^{\beta}_a \otimes \epsilon^{\alpha\beta}_{a} \otimes \theta^{\alpha\beta}_a\, .
\end{align}
\noindent All repeated indices are to be summed over. The $\eta_{\alpha}$ are the linearly independent constant spinors on $\mathbb{R}^3$, $\alpha= 1,2$, and the $\chi^{\beta}$ are the two sets of linearly independent Killing spinors on $S^4$, $\beta=1,2,3,4$, which can be taken to satisfy
\begin{align}
\tilde{\nabla}_b \chi_a^{\beta} = \frac{a}{2} \gamma_{S^4} \gamma_b \chi^{\beta}_{a} \hspace{0.7in} \gamma_{S^4} \chi_a^{\beta} = \chi_{-a}^{\beta}\;
\end{align}
\noindent where we hope it is clear that the $a=\pm$ appearing here is not a spacetime index on $S^4$ but rather a label of the two different signs in the Killing spinor equation.  The $\epsilon_a^{\alpha\beta}$ are commuting spinors on $\mathcal{M}_3$, and the $\theta_a^{\alpha\beta}$ are two-component spinors. The chirality condition (\ref{chirality}) implies that $\sigma^3 \theta_a^{\alpha\beta} = -\theta_a^{\alpha\beta}$, so that without loss of generality we can set 

$$
\theta_a^{\alpha\beta} = \begin{pmatrix}
0\\1
\end{pmatrix} \quad \forall\, a,\alpha ,\beta \, .
$$

Following \cite{deg}, we also note that, again without loss of generality, we can impose a reality condition on the basis Killing spinors\footnote{We thank John Estes for pointing this out.}. Specifically, we impose\footnote{It is impossible to impose $b_3 \eta^*=\eta$ on our basis, since $b_3^* b_3=-1$ (there are no Majorana spinors in three Euclidean dimensions.) Furthermore we see that we cannot impose $\gamma_{S^4} \chi_a = \chi_{-a}$ and $b_4 \chi_a=\chi_{-a}$ simultaneously, since $(\gamma_{S^4} b_4)^* (\gamma_{S^4} b_4) = -1$. However, we can impose a reality condition on the whole basis (rather than each factor indivually), since $(b_3 \gamma_{S^4} b_4)^* (b_3 \gamma_{S^4} b_4) = 1$. }
\bea
(b_3 \otimes b_4) (\eta^* \otimes \chi_a^*) &=& \eta \otimes \chi_{-a}
\eea

We can now reduce the BPS to three dimensions, by writing them in terms of the two complex two-component spinors $\epsilon_{\pm}$ on $\mathcal{M}_3$:
\bea
\label{kse2d}
2 \slashed{P}\, \epsilon_{-a}^* - \frac{1}{2}\e^{-3A} g \, \epsilon_{-a} + \frac{1}{2} h \, \epsilon_a &=& 0\\
\label{kse2i}
\frac{i}{2} \slashed{\partial} A \epsilon_{-a} + \frac{1}{2} \e^{-4B}  \slashed{\partial} f \epsilon_a - \frac{i}{16} (3 \e^{-3A} g) \epsilon_{-a}^*  - \frac{i}{16} h \epsilon_a^* &=& 0\\
\label{kse2a}
-\frac{a}{2} \e^{-B} \epsilon_{-a} + \frac{i}{2} \slashed{\partial} B \epsilon_a - \frac{1}{2} \e^{-4B} \slashed{\partial} f \epsilon_{-a} + \frac{i}{16} (\e^{-3A} g) \epsilon_a^* - \frac{i}{16} h \epsilon_{-a}^* &=& 0\\
\label{kse2mu}
\mathrm{D}_{\mu}\epsilon_a + \frac{i}{2} \e^{-4B} \slashed{\partial} f \gamma_{\mu} \epsilon_{-a} + \frac{1}{16} (\e^{-3A} g)\gamma_{\mu} \epsilon_a^* + \frac{1}{16} 3h \gamma_{\mu} \epsilon_{-a}^* &=& 0
\eea
Since no operators which affect the $\alpha,\beta,\cdots$ indices appear in (\ref{kse1}), these indices can be omitted, with the understanding that there is a $2\times 4 = 8$-fold multiplicity in each set of solutions $\{\epsilon_{+},\epsilon_{-}\}$ we shall find of (\ref{kse2d})-(\ref{kse2mu}). 

It is convenient to introduce a `tau-matrix' notation for these equations, as follows:
\begin{align}
(\tau^I \epsilon)_a \equiv \tau^I_{ab} \epsilon_b\quad I=0,1,2,3
\end{align}
where $\tau^{1,2,3}$ are the usual Pauli matrices acting on the $a,b$ indices, and $\tau^0\equiv \mathds{1}^{2\times 2}_{ab}$. 
Now the BPS equations reduced to $(2+1)$D read
\begin{align}
\slashed{P} \epsilon^* - \frac{1}{4} (\e^{-3A} g - h \tau^1 )\epsilon &= 0\label{d}\tag{d}\\
\label{i}\tag{i}
\slashed{\partial} A \epsilon - i \e^{-4B}  \slashed{\partial} f \tau^1 \epsilon - \frac{1}{8} (3 \e^{-3A} g + h \tau^1)\epsilon^* &= 0\\
\label{a}\tag{a}
-\e^{-B} \tau^2 \epsilon + \slashed{\partial} B \epsilon + i \e^{-4B} \slashed{\partial} f \tau^1 \epsilon + \frac{1}{8} (\e^{-3A} g - h \tau^1 )\epsilon^* &= 0\\
\label{mu}\tag{$\mu$}
\nabla_{\mu}\epsilon -\frac i2 Q_{\mu}\epsilon + \frac{i}{2} \e^{-4B} \slashed{\partial} f \gamma_{\mu} \tau^1 \epsilon + \frac{1}{16} (\e^{-3A} g +  3h \tau^1 ) \gamma_{\mu} \epsilon^* &= 0
\end{align}
%
%
%
%
%
%
%
%
%
\section{Spinor bilinear analysis}
\label{app:b}

We now solve the BPS system (\ref{d})-(\ref{mu}) using the standard techniques of bilinear analysis. 
First we introduce the real bilinears
\bea
f^{(I)} \equiv \epsilon^{\dagger} \sigma^2 \tau^I \epsilon&&V_{\mu}^{(I)} \equiv i\, \epsilon^{\dagger} \sigma^2 \tau^I \gamma_{\mu} \epsilon\quad .
\eea
Likewise we have the complex bilinears 
\bea
\tilde{f}^{(I)} \equiv \epsilon^t \sigma^2 \tau^I \epsilon&&\tilde{V}^{(I)}_{\mu} \equiv \epsilon^{t} \sigma^2 \tau^I\gamma_{\mu} \epsilon\quad .
\eea
Note that $\tilde{V}_{(2)}$ and $\tilde{f}^{(0,1,3)}$ all vanish identically since they are of the form $\epsilon^t M\epsilon$ where $M$ is an antisymmetric matrix, and we have taken $\epsilon$ to be commuting. For typographical clarity, we use the $(I)$ symbols as both subscripts and superscripts, but we intend no difference in meaning. The real bilinears have $q=0$, and the complex ones have $q=1$. 
We split the analysis into two, as is typical: 
\begin{itemize}
\item On the one hand we have algebraic equations among the bilinears implied by the BPS equations. We will use these to define a preferred orthonormal basis for the tangent space of $\mathcal{M}_3$, namely an identity structure, and we express the fluxes in terms of this. 
\item On the other hand there are differential equations which give the `torsion' of the identity structure, and which we use to define local coordinates and a metric. 
\end{itemize}

At various points we will use the 3D Fierz identities, which express linear dependence between the bilinears.
\subsection{Algebraic constraints}\label{algebra}
The first step is to the reduce the BPS equations to conditions on the minimum number of bilinears. We look at $\epsilon^{\dagger} \sigma^2$[(\ref{i})+(\ref{a})] and take real and imaginary parts to find that $f^{(2)}=0$ and $V^{(0)}\cdot\partial B=0$. 
Now taking $\epsilon^{\dagger} \tau^{0,1} \sigma^2$ \{(\ref{i}),(\ref{d})*\}, we find
\begin{align}\label{vanishing}
V^{(0)}\cdot X &= 0 \qquad X=\dd A,\, \dd B,\, \dd f,\, P
\end{align}
Next $\epsilon^{\dagger} \sigma^2\tau^2\gamma_{\mu}$[(\ref{i})-(\ref{a})]$+\text{c.c.}$ gives
$$
-4\,\e^{-4B}\dd f\,f^{(3)}=0
$$
which is only solved for $f^{(3)}=0$, assuming $\dd f \neq 0$. Since $\dd f = 0$ would lead to solutions preserving more supersymmetry than the 8 SUSYs we are interested in (for example the D1-brane solution), we ignore this possibility. 
Finally, using Fierz identities we can now show that 
\bea
f^{(0)}=0&&V^{(1)}=0\\ V_{(2,3)}^2 = -V_{(0)}^2 = f_{(1)}^2 &&V^{(I)} \cdot V^{(J)} = 0\quad\forall\; I\neq J\quad .
\eea
Now we have simplified things considerably. In particular we see that
\be
\{\frac{1}{f^{(1)}}V_{(0)},\frac{1}{f^{(1)}}V_{(3)}, \frac{1}{f^{(1)}}V_{(2)}\}\equiv \{e^0,e^1,e^2\}\nonumber
\ee
form an orthonormal basis for the cotangent space (and by raising indices, for the tangent space), and so we can take them as the vielbeine on $\mathcal{M}_3$. 
Thus the mininum set of bilinears to consider consists of one real scalar, one complex scalar, and three real vectors: $\left\{f_{(1)}, \tilde{f}_{(2)}, V^{(0)}, V^{(2)}, V^{(3)}\right\}$. We next find expressions for the fluxes in terms of them. 
\subsubsection*{Complex 3-form}
Since $\tilde{f}^{(2)}$ is the only complex bilinear, we expect its phase to control the phases of $h,g$ and $P$. Taking $\epsilon^{\dagger} \tau^{3,2} \sigma^2$(\ref{d})* and solving for $g$ and $h$, we find that
\begin{align}
h&=\frac{4}{\tilde{f}^{(2)}} V^{(3)} \cdot P\label{h}\\
\e^{-3A}g&=\frac{4i}{\tilde{f}^{(2)}} V^{(2)} \cdot P\label{g} \quad .
\end{align}
\subsubsection*{Five-form}
This can be obtained by taking $\epsilon^{\dagger} \sigma^2\gamma_{\mu}$(\ref{i})$+$(\ref{i})$^{\dagger} \sigma^2 \gamma_{\mu}\epsilon$:
\begin{align}
-2i\,f^{(1)}\,\e^{-4B} \partial_{\mu} f + 2 \partial_{\nu} A \epsilon^{\dagger} \sigma^2 \gamma_{\mu}^{\;\nu} \epsilon + \frac{1}{8} \left( (3 \e^{-3A} g \tilde{V}_{\mu}^{(0)*} + h \tilde{V}_{\mu}^{(1)*}) - \text{h.c} \right) = 0
\end{align}
Completeness of the tangent space implies that the $\tilde{V}^{(I)}$ are linear combinations of the other vector bilinears, and indeed a Fierzing gives $\tilde{V}^{(0)} = \frac{\tilde{f}^{(2)}}{f^{(1)}} V^{(3)}$, $\tilde{V}^{(1)} = i\frac{\tilde{f}^{(2)}}{f^{(1)}} V^{(2)}$. This leads to
\begin{align}
f_{(1)}^2\,\e^{-4B} \dd f = \left[V^{(2)}\cdot \left(\dd A + \frac{3}{2} \e^{-2i\theta} P\right)\right]\, V^{(3)} - \left[V^{(3)}\cdot \left(\dd A + \frac{1}{2} \e^{-2i\theta} P\right)\right]\, V^{(2)}\label{df}
\end{align}
where we have defined $\e^{i\theta}$ to be the phase of $\tilde{f}_{(2)}$. 

\subsubsection*{Axiodilaton}
We take the three combinations $\epsilon^{\dagger} \sigma^2 \tau^{2,3,0}$((\ref{i})+(\ref{a})). Plugging in the expressions for the fluxes obtained above we get
\begin{equation}
\begin{aligned}\label{combs}
V^{(0)}\cdot \left[\partial (A+B)+\e^{-2i\theta} P\right] &= 0\\
V^{(2)}\cdot \left[\partial (A+B)+\e^{-2i\theta} P\right] &= 0\\
V^{(3)}\cdot \left[\partial (A+B)+\e^{-2i\theta} P\right] &= \e^{-B} f^{(1)}
\end{aligned}
\end{equation}
Using the orthonormality of our tangent space basis, the above equations imply that
\begin{align}
\e^{-2i\theta} P = \frac{\e^{-B}}{f^{(1)}} V^{(3)}-\dd(A+B) \qquad .\label{p}
\end{align}
We can therefore see that $\e^{-2i\theta} P$ is real (this is discussed in section \ref{reality}). This implies that $P=\e^{2i\theta} \tilde P$, where $\tilde P$ is a real one-form. 

As a last piece of information to take from the algebraic conditions, we take $\epsilon^t\sigma^2[(\ref{i})+(\ref{a})]$ and use the expressions for the fluxes to find that $|\tilde{f}_{(2)}|^2=f_{(1)}^2$, so we can write
\begin{align}
\tilde{f}^{(2)} = \e^{i\theta}\,f^{(1)}\quad . \label{phase}
\end{align}
In summary, we have defined an identity structure, found the fluxes \eqref{h},\eqref{g},\eqref{df}, \eqref{p} in terms of it, and obtained the relation (\ref{phase}). 
\subsection{Torsion}\label{torsion}
In section \ref{algebra} we reduced the problem of solving the BPS equations to finding two scalars $\{f_{(1)}, \tilde{f}_{(2)}\}$ and three vector bilinears $\{V^{(0)}, V^{(2)}, V^{(3)}\}$. These satisfy a system of differential equations which is implied by the BPS equations:
\begin{align}
\dd (\e^{A+2B}f^{(1)}) &= 2\,\e^{A+B} V^{(3)}\label{f1diff}\\
\mathrm{D} (\e^{A+2B}\tilde{f}^{(2)}) &=2\,\e^{A+B} \tilde{V}^{(0)}\label{ft2diff}\\
\dd (\e^{2A+4B} V^{(0)}) &= - 4\,\e^{2A+3B} \ast V^{(2)}\label{v0diff}\\
\dd (\e^{2A+4B}V^{(2)}) &= -2\,\e^{2A} \dd f \wedge V^{(3)} - 4\,\e^{2A+3B} \ast V^{(0)}\label{v2diff}\\
\dd (\e^{2A+4B}V^{(3)}) &= 2\,\e^{-4B} \dd f \wedge V^{(2)}\label{v3diff}
\end{align}
We begin by showing that $V^{(0)}$ is a timelike Killing vector. It is timelike since $V_{(0)}^2 = - f_{(1)}^2$ is negative, and it satisfies
\begin{align}
\nabla_{(\mu} V^{(0)}_{\nu)} = -g_{\mu\nu} V^{(0)} \cdot \partial (A+2B) =0\quad ,
\end{align}
where the last equality follows from (\ref{vanishing}). Therefore $V^{(0)}$ satisfies the Killing equation on $\mathcal{M}_3$, and together with (\ref{vanishing}) this implies it is a Killing vector of the whole 10D metric. We define the coordinate $t$ such that $\partial / \partial t = V^{\#}_{(0)}$, where the notation denotes $V^{(0)}$ as a vector. 

We would like to find out about the phase $\e^{i\theta}$, which as we have already seen governs the phases of the complex fields ($g,h,P$). First we Fierz $\tilde{V}^{(0)}$, and then equate the LHSs of (\ref{ft2diff}) and (\ref{f1diff}). Together with (\ref{phase}) this gives $Q=\dd\theta$, so that the phase is just the $U(1)$ holonomy.

Next we consider the $\dd f^{(1)}$ equation (\ref{f1diff}). This implies that $\e^{A+B} V^{(3)}$ is a closed form, and defining a local coordinate $y^2 \equiv \e^{A+2B} f^{(1)}$ we have that
\begin{align}
V^{(3)} = y\, \e^{-(A+B)} \dd y\quad . 
\end{align}

Using (\ref{df}) and (\ref{combs}), the $\dd V^{(2)}$ equation (\ref{v2diff}) becomes
\begin{align}
\dd V^{(2)} = 3\,\left(\frac{\dd y}{y} - \, \partial_y (A+B) \dd y\right) \wedge V^{(2)}
\end{align}
so that we can write
\begin{align}
V^{(2)} = y^3 \e^{-3(A+B)} \dd x\qquad .
\end{align}
We take $x$ to be the final coordinate. We have automatically $\partial /\partial x \cdot \partial /\partial y = 0$. 

Lastly we turn to the equation for $\dd V^{(0)}$, which reads
\begin{align}
\dd V^{(0)} = (-2\,\dd (A+2B) +4\frac{\dd y}{y} ) \wedge V^{(0)}
\end{align}
so that
\begin{align}
V^{(0)} = y^4 \e^{-2A-4B} (\dd t + \omega)
\end{align}
for some closed form $\dd\omega = 0$. 

In summary, we can now write down the full form of the metric in terms of the warp factors $\e^A,\e^B$:
\begin{align}
\boxed{
\dd s_3^2 = - y^4 \e^{-2A-4B} (\dd t + \omega)^2 +\frac{\e^{2B}}{y^2} \dd y^2 + y^2\, \e^{-4A-2B} \dd x^2}
\end{align}
and the fluxes are 
\begin{subequations}
\begin{empheq}[box=\fbox]{align}
P &= \e^{2i\theta}\dd (\log y-A-B)\\
g&=-4i\,\e^{i\theta}\e^{5A+B}y^{-1}\partial_x (A+B)\\
h&=4\,\e^{i\theta}\e^{-B}y\,\partial_y (\log y-A-B)\\
\dd f &=\frac {y^4}{4}\left[ \partial_x (y^{-6}\e^{2A+6B})\, \dd y +\partial_y (y^{-2}\e^{-2A+2B})\, \dd x\right]
\end{empheq}
\end{subequations}
\subsection{Reality condition and $SL(2,\mathbb{R})$}\label{reality}
We now address the reality condition implied by (\ref{p}):
\begin{align}
\mathfrak{Im}(\e^{-2i\theta} P)=0
\end{align}
This implies that $P=\e^{2i\theta} \tilde P$, where $\tilde P$ is a real one-form. Then (\ref{pbianchi}) and (\ref{qfieldstrength}) imply $\dd Q=0$ and $Q=\dd\theta$, so that $Q$ is pure gauge (these relations are actually implied by the BPS equations, as shown in (\ref{torsion})). Therefore by a local $U(1)$ gauge transformation $\mathcal{U}=\e^{-i\theta}$, we can map to real $P$, i.e. vanishing axion. We know this must be the accompanying gauge transformation to an $SL(2,\mathbb{R})$ action, so the BPS equations give just the solution with $C^{(0)}=0$ and its orbit under S-duality, parametrized by $\theta$. This situation is familiar from e.g. \cite{deg}.

\section{Equations of motion}\label{eom}

In this appendix, we verify that our solution to the fermionic variations is also a solution to the equations of motion up to a unique Poisson type equation for the dilaton that has to be solved separately. Furthermore, our system is naturally equipped to include explicit sources in the form of fundamental strings and D3-branes smeared appropriately.

Since we have found that our configurations have $C^{(0)}=0$ up to a duality transformation, we set $\theta=0$ in this section, so that now $P=\frac{1}{2}\dd\phi$. We also set $\omega=0$.  In this case, it is easy to write the solution to the BPS equations in terms of $A$ and $\phi$ instead of $A$ and $B$:
\begin{align}\label{susyconf}
\dd s^2 = - \e^{2(A+\phi)} \dd t^2  +\e^{2A} \dd x^i \dd x^i + \e^{-2A} \left(\e^{-\phi} (\dd y^2 + y^2 \dd\Omega_4^2)+ \e^{\phi} \dd x^2 \right)\nonumber\\[2mm]
F_5 = \frac{y^4}{4} (1+\ast)\, \left(- \partial_x (\e^{-4A-3\phi})  \dd y+\partial_y(\e^{-4A-\phi}) \dd x \right)\wedge vol_{S^4}\\[2mm]
H_3 = \partial_y (\e^{2\phi})\,\dd t \wedge \dd y \wedge \dd x \qquad F_3 = 2\, \e^{4A-\phi}\partial_x \phi\,  \dd x^1\wedge\dd x^2\wedge\dd x^3\nonumber
\end{align}
%
%
%
We will now show in detail that all the equations of motion and Bianchi identities are satisfied provided we have $Q_B=-2\,\e^{4A}\partial_x\e^{-\phi}$, where $Q_B$ is a (non-zero) real constant, and the following equation, which we will refer to as the `Poisson equation', is satisfied:
\bea
\frac{1}{y^4}\partial_y (y^4\partial_y\e^{-2\phi})+\frac{1}{2}\partial_x^2\e^{-4\phi}=0\,. \label{poisson}
\eea
The equations of motion deriving from Type IIB supergravity, with a vanishing axion, together with the Bianchi identities, read
\begin{align}
&\dd (\e^{-\phi}\ast H_3)+4F_5\wedge F_3=0\,,\quad&\dd F_5-\frac{1}{4}H_3\wedge F_3=0\,,\nonumber\\[2mm]
&\dd (\e^{\phi} \ast F_3)+4 H_3\wedge F_5=0\,,&\dd F_3=0\,,\\[2mm]
&\dd \ast \dd \phi +\frac{1}{2}G_3\wedge\ast G_3 = 0\,.\nonumber
\end{align}
On top of that we have the Einstein equations
\begin{eqnarray}
R_{MN}&=&\frac{1}{2}\partial_M\phi \partial_N\phi\, +\frac{1}{6}(F_5)_{MP_1P_2P_3P_4}\, (F_5)_{N}^{\phantom{M}P_1P_2P_3P_4}\nonumber\\[2mm]
&+& \frac{1}{4}\mathfrak{Re}\left[(G_3)_{MP_1P_2}(G^*_3)_{N}^{\phantom{M}P_1P_2}\right]-\frac{1}{48}g_{MN}(G_3)_{P_1P_2P_3}(G^*_3)^{P_1P_2P_3}\,.\label{einsteineq}
\end{eqnarray}

\subsection*{$F_3$ Bianchi identity}

This states that $F_3=Q_B\, \dd x^1\wedge\dd x^2\wedge\dd x^3$ for some constant $Q_B$ related to the number of D5-branes, and thus to the number of baryon vertices. Using the BPS expression for $F_3$, we obtain $Q_B=-2\,\e^{4A}\partial_x\e^{-\phi}$, which is the advertised relationship between $A$ and $\phi$. 

\subsection*{$F_5$ Bianchi identity}

Substituting in the solutions of the BPS equations we obtain explicit expressions for $F_5$ 
\begin{align}\label{simplef5}
\dd f &= \frac{y^4}{4}\left[\partial_y(\e^{-4A-\phi}) \dd x - \partial_x (\e^{-4A-3\phi}) \dd y\right]\nonumber\\[2mm]
\Rightarrow \mathcal{F}_2=\e^{3A-4B}\ast_3 \dd f &= -\frac14\,\e^{8A+2\phi}\left(\partial_y (\e^{-4A-\phi}) \dd y +\e^{2\phi} \partial_x (\e^{-4A-3\phi}) \dd x\right)\wedge \dd t\\[2mm]
&=(\frac{1}{4}\dd (\e^{4A+\phi}) -\frac12  \e^{4A+2\phi}\partial_x(e^{-\phi})\, \dd x)\wedge \dd t\nonumber
\end{align}
The Bianchi identity comes in two pieces: a piece proportional to $vol_{S^4}$ and another proportional to $vol_{\mathbb{R}^3}$. We deal with the second part first. Using the $F_3$ Bianchi, we can express the two sides as follows: 
\begin{align}
\mathcal{F}_2&=\frac{1}{4}(\dd (\e^{4A+\phi}) + Q_B\,\e^{2\phi} \, \dd x)\wedge\dd t\nonumber\\[2mm]
\Rightarrow \dd F_5 &= \cdots +\frac{Q_B}{4}\partial_y \e^{2\phi} \dd y\wedge\dd x\wedge\dd t\wedge vol_{\mathbb{R}^3}\\[2mm]
\frac{i}{8}G\wedge G^*=\frac{1}{4}H_3\wedge F_3&=\frac{1}{4}\partial_y\e^{2\phi} \dd t\wedge\dd y\wedge\dd x\wedge(Q_B\,vol_{\mathbb{R}^3})=-\dd F_5\nonumber
\end{align}
so we see that this part is automatically satisfied. Now we turn to the first term. Setting this to zero amounts to saying we can locally find a function $f(x,y)$ whose derivative equals $\dd f$. Again using the $F_3$ Bianchi we have 
\begin{align}\label{simpledf}
\dd f = -\frac{y^4}{4\,Q_B}\left(\partial_y\partial_x \e^{-2\phi} \dd x - \frac{1}{2}\partial_x^2\e^{-4\phi} \dd y\right)\wedge vol_{S^4}\, .
\end{align}
The integrability condition $\dd^2 f=0$ gives
\begin{align}
\partial_x \left( \frac{1}{y^4}\partial_y (y^4\partial_y\e^{-2\phi})+\frac{1}{2}\partial_x^2\e^{-4\phi} \right)\, =\, 0\,,
\end{align}
which is verified if the Poisson equation is satisfied. Moreover, we can integrate (\ref{simpledf}) to obtain $f$. This equation states that
\begin{equation}
\partial_xf=\partial_x\left(-\frac{y^4}{4\,Q_B}\partial_y\e^{-2\phi}\right)\qquad\Rightarrow\qquad f=-\frac{y^4}{4\,Q_B}\partial_y\e^{-2\phi}+g(y)
\end{equation}
for some function $g(y)$. Using again (\ref{simpledf}) we have the compatibility condition
\begin{equation}
\partial_yf=-\frac{1}{4\,Q_B}\partial_y\left(y^4\partial_y\e^{-2\phi}\right)+g'(y)=\frac{1}{4Q_B}\,\frac12\,y^4\partial_x^2\e^{-4\phi}\,.
\end{equation}
Poisson's equation then states that $g(y)$ is at most a constant that we fix to zero, so we have
\begin{equation}\label{explicitf}
f=-\frac{y^4}{4\,Q_B}\partial_y\e^{-2\phi}\,,
\end{equation}
which is the expression used to find the brane charges.
\subsection*{$H_3$ equation of motion}

For the first term, we can write 
\begin{align}
\e^{-\phi}\ast H_3&= y^4 \,\partial_y(\e^{-2\phi})\, vol_{\mathbb{R}^3} \wedge vol_{S^4}\nonumber\\[2mm]
\Rightarrow \dd (\e^{-\phi}\ast H_3) &=\left[\partial_y (y^4 \partial_y(\e^{-2\phi})) \dd y +y^4 \partial_x\partial_y(\e^{-2\phi}) \dd x\right]\,\wedge vol_{\mathbb{R}^3} \wedge vol_{S^4}\,.
\end{align}
Analogously, for the the second term, using (\ref{simpledf}) we immediately have
\begin{align}
-4 F_5\wedge F_3 = y^4 \left(\partial_x\partial_y\e^{-2\phi} \dd x- \frac12\partial_x^2\e^{-4\phi} \dd y\right)\wedge vol_{\mathbb{R}^3} \wedge vol_{S^4}\,.
\end{align}
Finally we obtain
\begin{align}
\dd (\e^{-\phi}\ast H_3)+4 F_5\wedge F_3=&\left[\partial_y (y^4\partial_y\e^{-2\phi})+\frac{y^4}{2}\partial_x^2\e^{-4\phi}\right] \dd y\wedge vol_{\mathbb{R}^3} \wedge vol_{S^4} \,,
\end{align}
which is proportional to our Poisson equation. The equation of motion for $F_3$ works in a similar manner. 

\subsection*{Dilaton equation of motion}

This is
\bea
\dd\ast\dd\phi+\frac{1}{2}G_3\wedge\ast G_3=&\dfrac{\e^{2A+3\phi}}{2}\left( \frac{1}{y^4}\partial_y (y^4\partial_y\e^{-2\phi})+\frac{1}{2}\partial_x^2\e^{-4\phi} \right)\ast 1
\eea
recognizable again as (\ref{poisson}).

\subsection*{Einstein equations}

The different components of the Einstein equations \eqref{einsteineq}, in flat indices, read
\bea
E_{\hat{i}\hat{i}}&=&-\frac{\e^{2\phi+6A}}{Q_B y}\partial_x \left( \frac{1}{y^4}\partial_y (y^4\partial_y\e^{-2\phi})+\frac{1}{2}\partial_x^2\e^{-4\phi} \right)\,,\nonumber\\[2mm]
E_{\hat{a}\hat{a}}&=&0\,,\\[2mm]
E_{\hat{0}\hat{0}}&=& -\frac {\e^{3\phi+2A}}{2} \left( \frac{1}{y^4}\partial_y (y^4\partial_y\e^{-2\phi})+\frac{1}{2}\partial_x^2\e^{-4\phi} \right)\,.\nonumber
\eea
We conclude that, once the Poisson equation (\ref{poisson}) is verified, not only (\ref{susyconf}) is a supersymmetric configuration of Type IIB supergravity preserving 1/4 of the supercharges, but is additionally a solution to the equations of motion.
\subsection{Sourcing}\label{sourcing}

In this setup it is straightforward to include an explicit source on the RHS of the Poisson equation
\begin{equation}\label{sourced}
\frac{1}{y^4}\partial_y (y^4\partial_y\e^{-2\phi})+\frac{1}{2}\partial_x^2\e^{-4\phi}=\rho(x,y)\,.
\end{equation}
In order to do so one needs to carefully smear distributions of both fundamental strings and D3-branes and let them backreact on the geometry. Of course, this changes the form of the equations of motion, including the appearance of new terms in the energy momentum tensor. The details as well as the conditions required to preserve supersymmetry in the process can be found in the appendix of \cite{Faedo:2013aoa}. In that case we studied distributions depending only on the coordinate $y$, but it is straightforward to incorporate the $x$-dependence. 

In the notation of \cite{Faedo:2013aoa}, we have now the smearing forms
\begin{eqnarray}
\Omega_8&=&-y^4\rho(x,y)\,\dd y\wedge vol_{\mathbb{R}^3} \wedge vol_{S^4}\,,\nonumber\\[2mm]
\Omega_6&=& y^4\rho_{\rm D3}(x,y)\,\dd y\wedge \dd x \wedge vol_{S^4}\,.
\end{eqnarray}
From these we can read the directions along which the strings and the branes are distributed. They are supplemented by the calibration forms
\begin{eqnarray}
\mathcal{K}_2&=&-\e^{\frac32\phi}\,\dd t\wedge\dd x\,,\nonumber\\[2mm]
\mathcal{K}_4&=&\e^{4A+\phi}\,\dd t\wedge vol_{\mathbb{R}^3}\,,
\end{eqnarray}
that verify the calibration conditions stated in \cite{Faedo:2013aoa}, ensuring the supersymmetry of the configuration. It can be checked that all the equations of motion, modified by the presence of the sources, are verified given the sourced Poisson equation (\ref{sourced}) and the condition
\begin{equation}
Q_B\,\rho_{\rm D3}\,=\,\partial_x\rho\,.
\end{equation}
This connection between the distribution of D3-branes and strings is also responsible for the preservation of the relation among the charges, and is natural given that every string has to end on a D3-brane.


\begin{thebibliography}{99}

\bibitem{grinstein} G. Grinstein, ``Anisotropic sine-Gordon model and infinite-order phase transitions in three dimensions,'' Phys.\ Rev.\ B {\bf 23}, 4615 (1981).

\bibitem{hornreich} R. Hornreich, M. Luban, and S. Shtrikman, Phys.\ Rev. \ Lett. {\bf 35}, 1678 (1975).

\bibitem{sachdev} S. Sachdev, ``Quantum Phase Transitions,'' Cambridge University Press, 1999.

\bibitem{fradkin} E. Fradkin, ``Field Theories of Condensed Matter Physics'',  Cambridge University Press, 2013.  


\bibitem{maldacena} 
  J.~M.~Maldacena,
  ``The Large N limit of superconformal field theories and supergravity,''
  Int.\ J.\ Theor.\ Phys.\  {\bf 38}, 1113 (1999)
  [Adv.\ Theor.\ Math.\ Phys.\  {\bf 2}, 231 (1998)]
  [hep-th/9711200].
  
\bibitem{witten} 
  E.~Witten,
  ``Anti-de Sitter space and holography,''
  Adv.\ Theor.\ Math.\ Phys.\  {\bf 2}, 253 (1998)
  [hep-th/9802150].

\bibitem{Kachru:2008yh} 
  S.~Kachru, X.~Liu and M.~Mulligan,
  ``Gravity duals of Lifshitz-like fixed points,''
  Phys.\ Rev.\ D {\bf 78}, 106005 (2008)
  [arXiv:0808.1725 [hep-th]].
  
  
\bibitem{Taylor:2008tg} 
  M.~Taylor,
  ``Non-relativistic holography,''
  arXiv:0812.0530 [hep-th].

\bibitem{Ogawa:2011bz} 
  N.~Ogawa, T.~Takayanagi and T.~Ugajin,
  ``Holographic Fermi Surfaces and Entanglement Entropy,''
  JHEP {\bf 1201}, 125 (2012)
  [arXiv:1111.1023 [hep-th]].
  
\bibitem{Huijse:2011ef}
  L.~Huijse, S.~Sachdev and B.~Swingle,
  ``Hidden Fermi surfaces in compressible states of gauge-gravity duality,''
  Phys.\ Rev.\ B {\bf 85} (2012) 035121
  [arXiv:1112.0573 [cond-mat.str-el]].

\bibitem{Dong:2012se} 
  X.~Dong, S.~Harrison, S.~Kachru, G.~Torroba and H.~Wang,
  ``Aspects of holography for theories with hyperscaling violation,''
  JHEP {\bf 1206}, 041 (2012)
  [arXiv:1201.1905 [hep-th]].

\bibitem{Witten:1998xy} 
  E.~Witten,
  ``Baryons and branes in anti-de Sitter space,''
  JHEP {\bf 9807}, 006 (1998)
  [hep-th/9805112].


\bibitem{Faedo:2013aoa} 
  A.~F.~Faedo, B.~Fraser and S.~P.~Kumar,
  ``Supersymmetric Lifshitz-like backgrounds from $\mathcal{N}$ = 4 SYM with heavy quark density,''
  JHEP {\bf 1402}, 066 (2014)
  [arXiv:1310.0206 [hep-th]].
  
  
\bibitem{Rajagopal:2000wf} 
  K.~Rajagopal and F.~Wilczek,
  ``The Condensed matter physics of QCD,''
  In *Shifman, M. (ed.): At the frontier of particle physics, vol. 3* 2061-2151
  [hep-ph/0011333].
  
  
\bibitem{Fukushima:2010bq} 
  K.~Fukushima and T.~Hatsuda,
  ``The phase diagram of dense QCD,''
  Rept.\ Prog.\ Phys.\  {\bf 74}, 014001 (2011)
  [arXiv:1005.4814 [hep-ph]].

\bibitem{CasalderreySolana:2011us} 
  J.~Casalderrey-Solana, H.~Liu, D.~Mateos, K.~Rajagopal and U.~A.~Wiedemann,
  ``Gauge/String Duality, Hot QCD and Heavy Ion Collisions,''
  arXiv:1101.0618 [hep-th].

\bibitem{Hartnoll:2009ns} 
  S.~A.~Hartnoll, J.~Polchinski, E.~Silverstein and D.~Tong,
  ``Towards strange metallic holography,''
  JHEP {\bf 1004}, 120 (2010)
  [arXiv:0912.1061 [hep-th]].


\bibitem{Bigazzi:2013jqa} 
  F.~Bigazzi, A.~L.~Cotrone and J.~Tarrio,
  ``Charged D3-D7 plasmas: novel solutions, extremality and stability issues,''
  JHEP {\bf 1307}, 074 (2013)
  [arXiv:1304.4802 [hep-th]].
  
\bibitem{Korchemsky:1991zp} 
  G.~P.~Korchemsky and A.~V.~Radyushkin,
  ``Infrared factorization, Wilson lines and the heavy quark limit,''
  Phys.\ Lett.\ B {\bf 279}, 359 (1992)
  [hep-ph/9203222].

\bibitem{Kumar:2012ui} 
  S.~P.~Kumar,
  ``Heavy quark density in N=4 SYM: from hedgehog to Lifshitz spacetimes,''
  JHEP {\bf 1208}, 155 (2012)
  [arXiv:1206.5140 [hep-th]].

\bibitem{Maldacena:1998im} 
  J.~M.~Maldacena,
  ``Wilson loops in large N field theories,''
  Phys.\ Rev.\ Lett.\  {\bf 80}, 4859 (1998)
  [hep-th/9803002].
  
\bibitem{Rey:1998ik} 
  S.~J.~Rey and J.~T.~Yee,
  ``Macroscopic strings as heavy quarks in large N gauge theory and anti-de Sitter supergravity,''
  Eur.\ Phys.\ J.\ C {\bf 22}, 379 (2001)
  [hep-th/9803001].

\bibitem{Azeyanagi:2009pr} 
  T.~Azeyanagi, W.~Li and T.~Takayanagi,
  ``On String Theory Duals of Lifshitz-like Fixed Points,''
  JHEP {\bf 0906}, 084 (2009)
  [arXiv:0905.0688 [hep-th]].


\bibitem{Faedo:2014ana} 
  A.~F.~Faedo, A.~Kundu, D.~Mateos and J.~Tarrio,
  ``(Super)Yang-Mills at Finite Heavy-Quark Density,''
  arXiv:1410.4466 [hep-th].



  
\bibitem{Drukker:2005kx} 
  N.~Drukker and B.~Fiol,
  ``All-genus calculation of Wilson loops using D-branes,''
  JHEP {\bf 0502}, 010 (2005)
  [hep-th/0501109].
  
  
\bibitem{yamaguchi} 
  S.~Yamaguchi,
  ``Wilson loops of anti-symmetric representation and D5-branes,''
  JHEP {\bf 0605}, 037 (2006)
  [hep-th/0603208].

\bibitem{Gomis:2006sb} 
  J.~Gomis and F.~Passerini,
  ``Holographic Wilson Loops,''
  JHEP {\bf 0608}, 074 (2006)
  [hep-th/0604007].

\bibitem{Hartnoll:2006is} 
  S.~A.~Hartnoll and S.~P.~Kumar,
  ``Higher rank Wilson loops from a matrix model,''
  JHEP {\bf 0608}, 026 (2006)
  [hep-th/0605027].

\bibitem{yamaguchi1} 
  S.~Yamaguchi,
  ``Bubbling geometries for half BPS Wilson lines,''
  Int.\ J.\ Mod.\ Phys.\ A {\bf 22}, 1353 (2007)
  [hep-th/0601089].
  
\bibitem{lunin} 
  O.~Lunin,
  ``On gravitational description of Wilson lines,''
  JHEP {\bf 0606}, 026 (2006)
  [hep-th/0604133].

  
\bibitem{deg} 
  E.~D'Hoker, J.~Estes and M.~Gutperle,
  ``Gravity duals of half-BPS Wilson loops,''
  JHEP {\bf 0706}, 063 (2007)
  [arXiv:0705.1004 [hep-th]].


\bibitem{gauntlett11d} 
  J.~P.~Gauntlett, J.~B.~Gutowski, C.~M.~Hull, S.~Pakis and H.~S.~Reall,
  ``All supersymmetric solutions of minimal supergravity in five- dimensions,''
  Class.\ Quant.\ Grav.\  {\bf 20}, 4587 (2003)
  [hep-th/0209114].
  
  
\bibitem{gutowski} 
  J.~P.~Gauntlett, J.~B.~Gutowski and S.~Pakis,
  ``The Geometry of D = 11 null Killing spinors,''
  JHEP {\bf 0312}, 049 (2003)
  [hep-th/0311112].
  
  
\bibitem{gmsw} 
  J.~P.~Gauntlett, D.~Martelli, J.~Sparks and D.~Waldram,
  ``Supersymmetric AdS(5) solutions of M theory,''
  Class.\ Quant.\ Grav.\  {\bf 21}, 4335 (2004)
  [hep-th/0402153].



\bibitem{deg1} 
  E.~D'Hoker, J.~Estes and M.~Gutperle,
  ``Exact half-BPS Type IIB interface solutions. I. Local solution and supersymmetric Janus,''
  JHEP {\bf 0706}, 021 (2007)
  [arXiv:0705.0022 [hep-th]].
  
  
\bibitem{hsingh} 
  H.~Singh,
  ``Lifshitz/Schr\'odinger Dp-branes and dynamical exponents,''
  JHEP {\bf 1207}, 082 (2012)
  [arXiv:1202.6533 [hep-th]];
  H.~Singh,
  ``Lifshitz to AdS flow with interpolating $p$-brane solutions,''
  JHEP {\bf 1308}, 097 (2013)
  [arXiv:1305.3784 [hep-th]].
  
  
\bibitem{deyroy} 
  P.~Dey and S.~Roy,
  ``Lifshitz-like space-time from intersecting branes in string/M theory,''
  JHEP {\bf 1206}, 129 (2012)
  [arXiv:1203.5381 [hep-th]];
  P.~Dey and S.~Roy,
  Phys.\ Rev.\ D {\bf 87}, no. 6, 066001 (2013)
  [arXiv:1208.1820 [hep-th]].
  
\bibitem{maldacenagaiotto} 
  D.~Gaiotto and J.~Maldacena,
  ``The Gravity duals of N=2 superconformal field theories,''
  JHEP {\bf 1210}, 189 (2012)
  [arXiv:0904.4466 [hep-th]].
  
  
\bibitem{llm} 
  H.~Lin, O.~Lunin and J.~M.~Maldacena,
  ``Bubbling AdS space and 1/2 BPS geometries,''
  JHEP {\bf 0410}, 025 (2004)
  [hep-th/0409174].

\bibitem{Headrick:2007ca} 
  M.~Headrick,
  ``Hedgehog black holes and the Polyakov loop at strong coupling,''
  Phys.\ Rev.\ D {\bf 77}, 105017 (2008)
  [arXiv:0712.4155 [hep-th]].

\bibitem{benini} 
  F.~Benini, F.~Canoura, S.~Cremonesi, C.~Nunez and A.~V.~Ramallo,
  ``Unquenched flavors in the Klebanov-Witten model,''
  JHEP {\bf 0702}, 090 (2007)
  [hep-th/0612118];
  F.~Benini, F.~Canoura, S.~Cremonesi, C.~Nunez and A.~V.~Ramallo,
  JHEP {\bf 0709}, 109 (2007)
  [arXiv:0706.1238 [hep-th]].

\bibitem{carlos} 
  C.~Nunez, A.~Paredes and A.~V.~Ramallo,
  ``Unquenched Flavor in the Gauge/Gravity Correspondence,''
  Adv.\ High Energy Phys.\  {\bf 2010}, 196714 (2010)
  [arXiv:1002.1088 [hep-th]].
  
  
\bibitem{bigazzi1} 
  F.~Bigazzi, A.~L.~Cotrone, J.~Mas, D.~Mayerson and J.~Tarrio,
  ``D3-D7 Quark-Gluon Plasmas at Finite Baryon Density,''
  JHEP {\bf 1104}, 060 (2011)
  [arXiv:1101.3560 [hep-th]].


\bibitem{bigazzi2} 
  F.~Bigazzi, A.~L.~Cotrone and J.~Tarrio,
  ``Charged D3-D7 plasmas: novel solutions, extremality and stability issues,''
  JHEP {\bf 1307}, 074 (2013)
  [arXiv:1304.4802 [hep-th]].
 
\bibitem{Chen:2009kx} 
  H.~Y.~Chen, K.~Hashimoto and S.~Matsuura,
  ``Towards a Holographic Model of Color-Flavor Locking Phase,''
  JHEP {\bf 1002}, 104 (2010)
  [arXiv:0909.1296 [hep-th]].
  
\bibitem{obannon} 
  M.~Ammon, K.~Jensen, K.~Y.~Kim, J.~N.~Laia and A.~O'Bannon,
  ``Moduli Spaces of Cold Holographic Matter,''
  JHEP {\bf 1211}, 055 (2012)
  [arXiv:1208.3197 [hep-th]].
  
\end{thebibliography}
\end{document}